\documentclass[preprint,aps,prd,a4paper,showpacs,nofootinbib,preprintnumbers,superscriptaddress]{revtex4-1} %twocolumn
\usepackage{hyperref}
\usepackage{mathtools,amsfonts,amssymb}
\usepackage{amsmath,amsthm}
\usepackage{graphicx}
\usepackage{color}

\begin{document}
	\title{Viable Constraint on Scalar Field in  Scalar-Tensor Theory}
	
%---author 1
\author{Chao-Qiang Geng}
\email[Electronic address: ]{geng@phys.nthu.edu.tw}
\affiliation{School of Fundamental Physics and Mathematical Sciences\\Hangzhou Institute for Advanced Study, UCAS, Hangzhou 310024, China}
\affiliation{International Centre for Theoretical Physics Asia-Pacific, Beijing/Hangzhou, China}
\affiliation{Department of Physics,
	National Tsing Hua University, Hsinchu 300, Taiwan}
\affiliation{Physics Division,
	National Center for Theoretical Sciences, Hsinchu 300, Taiwan}
%---author 2
\author{Hao-Jui Kuan}
\email[Electronic address: ]{guanhauwzray@gmail.com}
\affiliation{Department of Physics,
	National Tsing Hua University, Hsinchu 300, Taiwan}
%---author 3
\author{Ling-Wei Luo}
\email[Electronic address: ]{lwluo@gate.sinica.edu.tw}
\affiliation{Institute of Physics, Academia Sinica, Taipei 11529, Taiwan}

\date{\today}
	
%%%%%%%%%%%%%%%%%%%%%%%%%%%%%%%%%%%%%%%%%%%%%%%%%%%%%%%%%
%-----------------------------------------------------------------------%
% Abstract
%-----------------------------------------------------------------------%
\begin{abstract}
The scalar-tensor theory can be formulated in both Jordan and Einstein frames, which are conformally related  together with a redefinition of the scalar field. 
As the solution to the equation of the scalar field in the Jordan frame does not have the one-to-one correspondence with that in the Einstein frame, we give a criterion along with some specific models to check if the scalar field in the Einstein frame is viable or not by confirming whether this  field is reversible back to the Jordan frame. We further show that the criterion in the first parameterized post-Newtonian approximation can be determined by the parameters of the osculating approximation of the coupling function in the Einstein frame and can be treated as a viable  constraint on any numerical study in the scalar-tensor scenario.
We also  demonstrate that the Brans-Dicke theory with an infinite constant parameter $\omega_{\text{BD}}$ is a counterexample of the equivalence between two conformal frames due to the violation of the viable constraint.
\end{abstract}

\maketitle

%%%%%%%%%%%%%%%%%%%%%%%%%%%%%%%%%%%%%%%%%%%%%%%%%%%%%%%%%%
\section{Introduction}
General Relativity (GR) is a great theory as its predictions agree well with  experiments. However, some
strong-field phenomena may deviate from those of GR.
Jordan proposed a scalar field which couples to the spacetime curvature to fulfill the hypothesis
about the variation of the constant of gravitation~\cite{Jordan, Jordan:1959eg}. 
Bergmann~\cite{Bergmann:1948} and 
Fierz~\cite{Fierz:1956zz} used the Kaluza-Klein theory~\cite{KK} to get an effective gravitational constant. 
Among various alternative gravity theories, the
scalar-tensor (ST) one is the most natural extension to GR since it can be
obtained by reducing a higher dimensional theory into a four dimensional one together
with a scalar field.

The most famous formulation  of the  ST theory was done  by
Brans and Dicke (BD)~\cite{Brans:1961sx, Dicke:1961gz}, in which GR is explicitly modified by  
introducing a scalar field variably to determine the universal Newtonian coupling.
Subsequently, the BD theory was  generalized by Bergmann~\cite{Bergmann:1968ve} and 
Wagoner~\cite{Wagoner:1970vr} to include a scalar potential,
which has been widely used  to realize  inflation 
and the late-time cosmic acceleration of the universe.
Damour and Esposito-Farèse~\cite{Damour:1992we} further extended it 
with couplings between gravity and  multi-scalar fields, in which,
besides the tensor gravitational field, one or more scalar fields are added 
in the gravitational sector with non-minimal couplings.

There are two formalisms of the ST theory. One is given in the so called Jordan frame, where 
the scalar field $\phi$ couples non-minimally to the Ricci scalar $R$ but not directly to  matter
with the kinetic term for the  scalar field involving  an arbitrary function $\omega(\phi)$. 
The other is in the Einstein frame, where the 
canonical scalar field $\varphi$ is introduced to minimally couple to the Ricci scalar $R^{\star}$.
The later one makes the field equations mathematically less complicated due to the separation of the 
second-order derivatives of the gravitational variables $g^{\star}_{\mu \nu}$ and $\varphi$, but  the
matter couplings depend on the scalar field. 
In order to have a frame transformation, 
we need to assume that each scalar field is a  functions of the other. 
In other words, we should have $\phi(\varphi)$
to rephrase the system from the Jordan frame into the Einstein one, 
and vice versa.
The connection between these two frames is through a Weyl (conformal) transformation 
by taking the Jordan frame metric $g_{\mu \nu}$ into the Einstein frame one $g^{\star}_{\mu \nu}$, 
together with a redefinition of the original scalar field $\phi$ into $\varphi$ to 
have its kinetic term being a canonical form.

The Einstein frame is convenient to perform the numerical  simulations and
transparent to discuss the cosmological applications.
However, 
the issues on the physical interpretation and equivalence of these two frames 
have been debated for a long time (e.g.~\cite{Faraoni:2006fx, Capozziello:2010sc}).
On the other hand, there has been much less attention  to the redefinition of 
the scalar field. In spite of many efforts to show that 
these two conformal frames are equivalent, in general there are mainly two problems 
from the redefinition of the scalar field, which are shown as follows:
\\
(i) The existence of the Einstein frame is 
determined by whether the 
scalar field $\phi$ can be described in terms of the new field $\varphi$.
Furthermore, we need to transform the system back to the Jordan frame.
These operations are based on  
the requirement of the inequalities, given by
\begin{subequations}\label{Eq:E-frame requirement}
\begin{align}
&\frac{d \phi}{d\varphi} \ne 0 
\quad\text{(Jordam to Einstein frame)}, \\[0.3em]
&\frac{d\varphi}{d\phi} \ne 0
\quad\text{(Einstein to Jordam frame)},
\end{align} 
\end{subequations}
which  are related to each other, Note that the \emph{non-zero} values of the derivatives  should also be \emph{finite}.
We call \eqref{Eq:E-frame requirement} as the \emph{derivative constraints}.
In practice, one often study the ST theories in the Einstein frame even though most of 
them do not perceive the inverse problem of the solution in the Jordan frame. 
In other words, they do not consider the possibility that their results may not 
be able to be transformed back into the Jordan frame. 
Since the Jordan frame is assumed to be the physical frame, 
those fields in this case do not bear any physical meanings. 
It is one of the purposes of the present work to give an essential criterion to see
if a solution in the Einstein frame can  uniquely correspond to a Jordan one.
It will be shown that \eqref{Eq:E-frame requirement} leads to the required condition.
\\
(ii) The coefficient of the kinetic term of $\varphi$ determines 
the value of $d\varphi/d\phi$ through the Weyl transformation.
However, the sign of the differential relation $d\varphi/d\phi$
can be either ``$+$'' or ``$-$'', which has been mentioned in \cite{Damour:2007uf} 
but without further discussions therein,
while many just consider the positive sign as in \cite{Fujii, Dicke:1961gz, Sperhake:2017itk}.
Therefore, starting from a given ST model in the Einstein frame, it usually corresponds to 
two different models in the Jordan (physical) frame. We will give an example on this issue.
In addition, the condition that the relation $d\varphi/d\phi$ is regular 
indicates that there is an irrelevant value of $\varphi$ by which the solution space of
the scalar fields in the Einstein frame is divided into two branches \cite{Geng}.
Conversely, the value of $\varphi$ can restrict the choice of the sign of the relation 
$d\varphi/d\phi$ in some cases. 
As a result, it arises the uniqueness problem of a model in the Jordan frame.

In this study, we explore the ST theory in the Jordan and Einstein frames by examining
the relations between them. In particular, we would like to find out the possible constraints 
on the theory by requiring the one-to-one correspondence for the physical quantities 
after the conformal transformation between the two frames.

The paper is organized as follows. 
In Sec.~II, we first write down the ST action. 
We then display the transformation between the Jordan and Einstein frames
and  demonstrate  that there is a constraint on 
the scalar transformation. 
In Sec.~III,  we take two specific models as examples to illustrate how the constraint
manifest itself as a criterion to check if the solution in the Einstein frame is viable or not. 
We present our conclusions in Sec.~IV.

%\end{section}
%\newpage
%%%%%%%%%%%%%%%%%%%%%%%%%%%%%%%%%%%%%%%%%%%%%%%%%%%%%%%%%%%%%%
\section{Scalar-Tensor theories}
\subsection{ Conformal Frame Freedom}
A general action of the ST theory with a 
 dimensionless single scalar field 
can be written as\footnote{The metric signature is $(- + + +)$, 
	while the Riemann tensor is given in terms of the Christoffel symbol by 
	$R^{\alpha}{}_{\beta \mu \nu} 
	\coloneqq \Gamma^{\alpha}{}_{\beta \nu ,\mu} - \Gamma^{\alpha}{}_{\beta \mu ,\nu} 
	+ \Gamma^{\alpha}{}_{\sigma \mu}\Gamma^{\sigma}{}_{\beta \nu} 
	- \Gamma^{\alpha}{}_{\sigma \nu}\Gamma^{\sigma}{}_{\beta \mu}$.}
%--------------------------------------------------%
\begin{align}\label{action}
S = \int d^{4}x\, \frac{\sqrt{-g}}{16\pi G} \bigg( F(\phi)R - \mathcal{B}(\phi)
    g^{\mu \nu} \partial_{\mu}\phi \partial_{\nu}\phi -U(\phi)\bigg)
    + S_{m}[\psi_{m}, e^{2\gamma(\phi)}g_{\mu \nu}],
\end{align}
where $g$ and $R$ associated with $g_{\mu \nu}$ respectively
stand for the determinant and Riemann scalar curvature, 
$F(\phi)$, $\mathcal{B}(\phi)$, $U(\phi)$ and $\gamma(\phi)$ represent functions of $\phi$
and $\psi_{m}$ denotes the non-gravitational fields. 
We note that $F(\phi)>0$ in the Jordan frame.
The corresponding equations of motion for $g_{\mu\nu}$ and $\phi$ are
\begin{subequations}\label{Eq:EoM of g and phi}
\begin{align}
R_{\mu\nu} = \frac{1}{F}
             \bigg[8\pi G \bigg(T_{\mu\nu} - \frac{1}{2}g_{\mu\nu}T\bigg)
               + \frac{1}{2}g_{\mu\nu}U 
               + \mathcal{B}\,\partial_{\mu}\phi \partial_{\nu}\phi\bigg], \label{Eq:EoM of g}
\end{align}
and
\begin{align}
\square \phi = \frac{1}{2\mathcal{B}}\bigg[-F_{\phi} R + U_{\phi} 
                   - \mathcal{B}_{\phi}\, g^{\mu \nu} \partial_{\mu}\phi \partial_{\nu}\phi
                   - 16 \pi G \alpha T \bigg],  \label{Eq:EoM of phi}
\end{align}
\end{subequations}
respectively, where the subscript $\phi$ denotes the partial derivative with respect to $\phi$.
Here, we have used the relation
\begin{align}
\frac{\delta S_{m}}{\delta \phi} \coloneqq  \sqrt{-g}\, \alpha T
\end{align}
where
\begin{align}\label{Eq:alpha defind by gamma}
\alpha &\coloneqq \gamma_{\phi},
\end{align}
and $T \coloneqq g^{\mu \nu}T_{\mu \nu}$ with the stress energy tensor, given by
\begin{align}
	T_{\mu \nu} \coloneqq \frac{-2}{\sqrt{-g}} \frac{\delta S_{m}}{\delta g^{\mu \nu}}. % = A^{2}T_{\mu\nu}.
\end{align}

%It has been shown that 
This form of the action is \emph{unchanged} under a group 
of field redefinitions through a Weyl transformation.
Specifically, with the new metric $g^{\star}_{\mu\nu}$ and 
scalar field $\varphi$, defined by
\begin{subequations}
\begin{align}
	&\phi = \phi(\varphi), \label{Eq:phi of varphi}\\
	&g_{\mu\nu} = e^{2\Gamma(\phi(\varphi))}g^{\star}_{\mu\nu},
\end{align}
\end{subequations}
along with the  coupling function $\Gamma$
 and necessary condition \eqref{Eq:E-frame requirement},
the action~\eqref{action} has the same form up to a boundary term~\cite{Jarv:2014laa, Flanagan:2004bz},
i.e.,
\begin{align}\label{Eq:8}
	S = \int d^{4}x\, \frac{\sqrt{-g^{\star}}}{16\pi G} \bigg( F^{\star}(\varphi)R^{\star}
	    - \mathcal{B}^{\star}(\varphi)g^{\star\mu \nu} 
	      \partial_{\mu}\varphi \partial_{\nu}\varphi - U^{\star}(\varphi)\bigg)
	+ S_{m}[\psi_{m}, e^{2\gamma^{\star}(\varphi)}g^{\star}_{\mu \nu}],
\end{align} 
through the definition of
\begin{align}\label{Eq:relation of phi and varphi}
e^{\Gamma}\big( 
\mathcal{B} - 6\, F\Gamma_{\phi}^{2} 
  - 6\, F_{\phi}\Gamma_{\phi} \big)^{1/2}\frac{d \phi}{d \varphi}
\coloneqq \pm\sqrt{\mathcal{B}^{\star}}.
\end{align}
The equation \eqref{Eq:relation of phi and varphi} 
yields that 
\begin{align}\label{Eq:general E-frame requirement}
\frac{d \phi}{d \varphi} =
\pm \sqrt{
\frac{  e^{-2\Gamma} \mathcal{B}^{\star}  }
{  \mathcal{B} - 6\, F\Gamma_{\phi}^{2} 
  - 6\, F_{\phi}\Gamma_{\phi}  }  }\,,
\end{align}
which should be non-zero and finite.
Here, the transformed functions of $F^{\star}(\varphi)$, $\mathcal{B}^{\star}(\varphi)$, 
$U^{\star}(\varphi)$ and $\gamma^{\star}(\varphi)$ are given by
\begin{subequations}
\begin{align}
  &F^{\star}(\varphi) 
    = e^{2\Gamma(\phi(\varphi))}F\big(\phi(\varphi)\big), \\
  &\mathcal{B}^{\star}(\varphi) 
    = e^{2\Gamma(\phi(\varphi))}\phi_{\varphi}^2(\varphi)
      \bigg( \mathcal{B}\big(\phi(\varphi)\big)
             - 6\, F\big(\phi(\varphi)\big)\Gamma_{\phi}^{2}\big(\phi(\varphi)\big)
	         - 6\, F_{\phi}\big(\phi(\varphi)\big)\Gamma_{\phi}\big(\phi(\varphi)\big)  \bigg), \\
  &U^{\star}(\varphi) 
    = e^{4\Gamma(\phi(\varphi))}U\big(\phi(\varphi)\big), \\
  &\gamma^{\star}(\varphi) 
    = \gamma\big(\phi(\varphi)\big) + \Gamma\big(\phi(\varphi)\big),
\end{align}
\end{subequations}
respectively. Two special frames of the action often used in the literature are:
%\begin{enumerate}
%\item

 The \emph{Jordan frame}, which is characterized by $\gamma =0$ in (\ref{action}). 
 In this frame,
      matter obeys the weak equivalent principle (WEP)~\cite{Brans:1961sx, Dicke:1961gz, Salgado:2005hx}, 
      meaning that freely falling objects follow the geodesics of the 
      Jordan frame metric~\cite{Fujii}. 
      But the shortcuts of this frame  violates the energy conditions. 
      However, it is not a negative kinetic energy to be problematic, 
      but an energy that is unbounded from below. 
      For instance, although a negative energy is usually associated with the instability 
      and  runaway solution, the Minkowski space is stable against 
      inhomogeneous perturbations in ST~\cite{Faraoni:2004is}. 
      Furthermore, a positive energy theorem has been shown to hold for 
      a special ST theory in the Jordan frame~\cite{Bertolami:1987wj}. 
      We regard the Jordan frame as the physical frame in this work.
		
%\item 
The \emph{Einstein frame}, which is characterized by $F^{\star} = 1$,
 $\mathcal{B}^{\star}= constant$
      and $\gamma^{\star} \ne 0$ in (\ref{Eq:8}). 
      By contrast to the Jordan one, the energy conditions are satisfied in this frame, 
      whereas the WEP is violated~\cite{Brans:2005ra}. 
			%Attempting to bear WEP, 
%\end{enumerate}

\subsection{Transformation between the Jordan and  Einstein Frames}

In  the Jordan frame, 
$\mathcal{B}(\phi) = \omega(\phi)/\phi$ and $\gamma(\phi) =0$ in (\ref{action}),
the action is given by
\begin{align}\label{JD action}
S = \int d^{4}x\, \frac{\sqrt{-g}}{16\pi G} \bigg( F(\phi)R - \frac{\omega(\phi)}{\phi} 
g^{\mu \nu} \partial_{\mu}\phi \partial_{\nu}\phi -U(\phi)\bigg)
+S_{m}[\psi_{m}, g_{\mu \nu}].
\end{align}
This action can be cast into a conformal frame by a Weyl transformation 
\begin{align}
g_{\mu\nu} = A^{2}(\phi)g^{\star}_{\mu\nu},
\end{align}
where $A(\phi) = e^{\Gamma(\phi)}$ is the coupling function.
%with a redefinition of the scalar field which will be addressed below.} 
%-----------------------------------%
%The transformation rule for the Christoffel symbols can be obtained by substituting the 
%transformed metric into the definition
%\begin{align}
%\Gamma^{\star \mu}{}_{\nu\lambda} 
%= \Gamma^{\mu}{}_{\nu\lambda} - \bigg( \partial_{\nu}(\ln A)\delta^{\mu}_{\lambda}
%+\partial_{\lambda}(\ln A)\delta^{\mu}_{\nu} -\partial^{\mu}(\ln A)g_{\nu\lambda} \bigg).
%\end{align}
%%%%%%%%%%%%%%%%%%%%%%%%%%%%%%%%%%%%%%%%%%%%%%%%%%%%%%%%%%%%%%
%{\color{red}Similarly, we can calculate the transformed scalar curvature and obtain that}
%\begin{align}
%R =A^{-2} \bigg(R^{\star}+6\,\square^{\star}(\ln A) - 6 g^{\star \mu\nu} \partial_{\mu}(\ln A^{-1})
%\partial_{\nu}(\ln A^{-1}) \bigg).
%\end{align}
We can now rewrite action \eqref{JD action} as
\begin{align}\label{casting into Ein}
S = 
%\int d^{4}x \frac{\sqrt{-g^{\star}}}{16\pi G}A^{4} &\bigg[ FA^{-2}\bigg(R^{\star} 
%+ 6\square^{\star}(\ln A^{-1})
%- 6g^{\star\mu\nu} \partial_{\mu}(\ln A^{-1}) \partial_{\nu}(\ln A^{-1}) \bigg)    \nonumber\\
%&- \frac{\omega}{\phi}A^{-2}g^{\star\mu \nu} \partial_{\mu}\phi \partial_{\nu}\phi -U(\phi)\bigg]
%+S_{m}[\psi_{m}, A^{2}g^{\star}_{\mu\nu}]    \nonumber\\
%\overset{FA^{2}=1}{=} 
\int d^{4}x\, \frac{\sqrt{-g^{\star}}}{16\pi G} \bigg(R^{\star} 
& - 6\,\square^{\star}(\ln A)
   - 6g^{\star\mu\nu}(-1)^{2} \partial_{\mu}(\ln A) \partial_{\nu}(\ln A)   \nonumber\\
& - \frac{\omega}{\phi}A^{2}g^{\star\mu \nu} \partial_{\mu}\phi \partial_{\nu}\phi -A^{4}U(\phi)\bigg) 
+S_{m}[\psi_{m}, A^{2}g^{\star}_{\mu\nu}],   
\end{align}
where
%at the second equality 
we have used the relation of $FA^{2} =1$ to obtain the action in the Einstein frame.
Consequently, 
by using
\begin{align}
\partial_{\mu}(\ln A) 
%= \partial_{\phi}(\ln A) \partial_{\mu} \phi 
= -\frac{1}{2F}F_{\phi}\partial_{\mu} \phi\,,
\end{align}
we can simplify \eqref{casting into Ein} to
%The second term in the integrant of~\eqref{casting into Ein} is a boundary term, 
%thus results in a constant in the action which is disposable. 
%Hence~\eqref{casting into Ein} can be further simplified as following
\begin{align}\label{Eq:conformal transformation}
S =  \int d^{4}x\, \frac{\sqrt{-g^{\star}}}{16\pi G} \bigg[ R^{\star}
    - 2g^{\star\mu\nu} 
        \bigg( \frac{3 F_{\phi}^{2}}{4F^{2}}
              + \frac{\omega}{2\phi F}\bigg)
        \partial_{\mu} \phi\partial_{\nu} \phi 
    - A^{4}U(\phi)\bigg] + S_{m}[\psi_{m}, A^{2}g^{\star}_{\mu\nu}]
\end{align}
up to a divergence term of $6\,\square^{\star}(\ln A^{-1})$.
Through the redefinition of the scalar field with
\begin{align}\label{trans relation}
\frac{d\varphi}{d\phi}
\coloneqq \pm \sqrt{\frac{3 F_{\phi}^{2}}{4F^{2}}
	+\frac{\omega}{2\phi F}},
\end{align}
we can obtain the derivative of the  \emph{canonical field} $\varphi = \varphi(\phi)$ 
in the Einstein frame.
The equation \eqref{trans relation} is important and 
should satisfy 
the derivative constraints in \eqref{Eq:E-frame requirement}.
It will be shown that the sign can determine the range of $\varphi$ 
in particular models in Sec.~\ref{Sec:Constraints on Specific Models}. 
As a result, \eqref{Eq:conformal transformation} reads as
\begin{align}\label{Eq:Einstein action}
S = \int d^{4}x\, \frac{\sqrt{-g^{\star}}}{16\pi G} \bigg(R^{\star} 
      - 2g^{\star\mu\nu}\partial_{\mu}\varphi \partial_{\nu}\varphi 
      - 4V(\varphi) \bigg) 
    + S_{m}[\psi_{m}, A^{2}g^{\star}_{\mu \nu}],
\end{align}
where $V(\varphi) \coloneqq A^{4}U(\phi)/4$. 
Obviously, the matter fields  couple non-minimally to the scalar field through $A^{2}(\phi)$ in
\eqref{Eq:Einstein action}.

%%%%%%%%%%%%%%%%%%%%%%%%%%%%%%%%%%%%%%%%%%%%%%%%%%%%%%%%%%%%%%
By variating~\eqref{Eq:Einstein action} with respect to
$g^{\star\mu\nu}$ and $\varphi$, we derive the equations of motion of 
the tensor and scalar fields to be
\begin{subequations}\label{Eq:EoM of g and varphi}
\begin{align}
&R^{\star}_{\mu \nu} = 8\pi G \bigg( T^{\star}_{\mu \nu} 
  - \frac{1}{2}T^{\star}g^{\star}_{\mu \nu}\bigg)
  + 2\partial_{\mu}\varphi \partial_{\nu}\varphi 
  + 2Vg^{\star}_{\mu \nu}, \label{Eq:EoM of metric} \\[0.3em]
&\square^{\star}\varphi 
  = -4\pi G \alpha(\varphi)T^{\star} + \frac{dV}{d\varphi}, \label{Eq:EoM of varphi}
\end{align}
\end{subequations}
respectively,  where $\alpha(\varphi) $ is a function of $\varphi$  in the Einstein frame, defined by
\begin{align}\label{Eq:alpha}
\alpha(\varphi) = \frac{d\ln A}{d\varphi}
= -\frac{F_{\phi}}{2F}\frac{d \phi}{d \varphi}\,,
\end{align}
%{\color{blue}as function of $\varphi$ defined in the Einstein frame,
which is closely related to the derivative constraints.
It can be checked that the equations of motion can be read by substituting 
$F=1$, $\mathcal{B}=2$, $\phi=\varphi$ and $U=4V$ as well as $\gamma=\ln A$ 
into \eqref{Eq:EoM of g}, \eqref{Eq:EoM of phi} and \eqref{Eq:alpha defind by gamma}.
%[Note that the logarithmic derivative of the coupling function $A$ {\color{red}in \eqref{Eq:alpha} }
%measures the coupling strength between the scalar field and matter, 
%which can be clearly seen from~\eqref{Eq:EoM of varphi}.]

Clearly, it is non-trivial that the equations of \eqref{Eq:EoM of g and varphi}
derived by variating~\eqref{Eq:Einstein action} with
$g^{\star\mu\nu}$ and $\varphi$ in the Einstein 
frame are equivalent to those from the variations of the action~\eqref{JD action}
with respect to $g^{\mu\nu}$ and $\phi$ in the Jordan frame. 
From the inverse function theorem, $\phi(\varphi)$ exists
%\begin{align}
%	\frac{\delta}{\delta \varphi} = \frac{d\phi}{d\varphi} \frac{\delta}{\delta \phi},
%\end{align}
%the equations of the scalar field in these two frames are equivalent 
as long as \eqref{trans relation} never vanishes, 
i.e., the derivative constraints.
%the following holds
Hence, the solution to the scalar equation in the Einstein frame must 
satisfy \eqref{Eq:E-frame requirement}.
Otherwise, it is not a solution to the scalar equation in the Jordan frame, 
which is identified as the physical equation of motion.

%%%%%%%%%%%%%%%%%%%%%%%%%%%%%%%%%%%%%%%%%%%%%%%%%%%%%%%%%%%%%%
There is another way to understand the derivative constraints.
It is clear that $U(\phi)$ is a physical  potential as it is defined in the  Jordan frame, 
whereas  $V(\varphi)$ is an auxiliary one. 
In general, if there is a potential $V(\varphi)$ in the Einstein frame,
one cannot claim that a solution $\varphi$ of~\eqref{Eq:EoM of varphi} is physical 
unless $U(\phi)= 4 A^{-4}(\phi)V(\varphi(\phi))$ exists and is unique.
It is apparent that $V$ can be a function of $\phi$ if and only if $\varphi$ is that of $\phi$.
As a result, the derivative constraints should hold.

From the derivative constraints, 
%\begin{align}\label{criterion}
%	\frac{d\varphi}{d\phi} \ne 0,
%\end{align}
%it should be either 
we can further show that
\begin{align}
	\frac{d\varphi}{d\phi} > 0 \text{\quad or\quad} \frac{d\varphi}{d\phi} < 0\,,
\end{align}
representing the choice of the sign
in the RHS of \eqref{trans relation}, which cannot 
%be switched 
cross the critical value of 
$d\varphi/d\phi = 0$ during the evolution of the scalar field~\cite{Jarv:2007iq}. 
Moreover, to bear the one-to-one correspondence between $F$ and $\phi$, 
 $F_{\phi}$ must be non-vanishing, so that  the sign of $F_{\phi}$ cannot flip either.
Under the condition of $F_{\phi}\ne 0$, the action~\eqref{JD action} can be rewritten 
into that in the Brans-Dicke-Bergmann-Wagoner (BDBW) theory~\cite{Clifton:2011jh}.

%%%%%%%%%%%%%%%%%%%%%%%%%%%%%%%%%%%%%%%%%%%%%%%%%%%%%%%%%%%%%%
\subsection{PPN Parameters in Scalar-Tensor Theories}
As discussed in the literature~\cite{Damour:1993hw,Damour:1995kt,Damour:1996ke, Jarv:2007iq, Jarv:2014laa},
the deviation from GR for a ST theory in the parameterized post-Newtonian (PPN) regime 
can be expressed in terms of the asymptotic value of 
$\alpha$ at spatial infinity as well as its successive derivatives. 

At the first order of PPN,
it has been shown that the coupling function is sufficient to be determined by 
the osculating approximation~\cite{Damour:2007uf}
\begin{align}\label{coupling func}
	\ln A=\alpha_{0}(\varphi -\varphi_{0})+ \frac{1}{2}\beta_{0}(\varphi-\varphi_{0})^{2}
\end{align}
and 
\begin{align}\label{Eq:eq of alpha}
  \alpha = \alpha_{0} + \beta_{0}(\varphi-\varphi_{0})
\end{align}
where $\varphi_{0}$ is the asymptotic value of $\varphi$ at spatial infinity and 
$\alpha_{0}$ and $\beta_{0}$ are constants,
defined as 
\begin{subequations}
	\begin{align}
	&\alpha_{0} \coloneqq \alpha ( \varphi_{0} ), \\
	&\beta_{0} \coloneqq \frac{d\alpha}{d\varphi}( \varphi_{0} ).
	\end{align}
\end{subequations}

Classically, for the massive scalar field, $\varphi_{0}$ should be determined by 
the ground state of $\varphi$, i.e., the global minimum of 
the potential $V(\varphi)$, which satisfies
\begin{align}\label{Eq: potential condtion 1}
	\lim_{r\rightarrow \infty} \frac{dV}{d\varphi}\bigg|_{\varphi=\varphi_{0}} =0
\end{align}
and
\begin{align}\label{Eq: potential condtion 2}
	\lim_{r\rightarrow \infty} \frac{d^2V}{d\varphi^2}\bigg|_{\varphi=\varphi_{0}} > 0.
\end{align}
It turns out that if one expresses  the potential in terms of the Taylor expansion 
around $\varphi=\varphi_{0}$, the coefficient of the linear term should vanish. 
To obtain the ground state naturally, it is convenient to consider a shifted field of
\begin{align}
	\varphi ' = \varphi - \varphi_{0},
\end{align}
which simplifies \eqref{coupling func} to be
\begin{align}
	\ln A = \alpha_{0} \varphi ' + \frac{1}{2}\beta_{0} \varphi^{\prime\,2}\,.
\end{align}
Subsequently, we obtain
\begin{align}\label{Eq:lnF}
\ln F =-2\alpha_{0}\varphi'- \beta_{0}\varphi^{\prime\,2}
\end{align}
through the help of the relation $FA^{2} =1$.
Note that $\alpha(\varphi) = 0$
with  $\alpha_{0} = \beta_{0} =0$ leads to the result in GR with the coupling function $A$
being identically unit, implying that the scalar field couples to the Riemann curvature minimally 
in the Jordan frame.
Consequently, the potential becomes a Taylor series of $\varphi '$ 
at $\varphi '=0$ without the linear term, given by
\begin{align}
V(\varphi') = V_0
+ \frac{1}{2!}\frac{d^2V}{d\varphi'^2}\bigg|_{\varphi'=0}\varphi^{\prime\,2} + \cdots.
\end{align}
where  $V_0 \coloneqq V(\varphi_0)$ is the minimum of the potential $V$,
 which gives no contribution to the equation of motion of $\varphi'$.

Furthermore, as a \emph{radiative} 
coordinate \cite{Blanchet:1986dk, Damour:1992we} can be constructed
in the Einstein frame with the metric to be the Minkowski one asymptotically, 
 one finds that $g_{\mu \nu}(r\to\infty) \approx A_{0}^{2}\eta_{\mu \nu}$
with $A_{0} \coloneqq A(\varphi_{0})$ in the Jordan frame. 
Within the conditions \eqref{Eq: potential condtion 1} 
and \eqref{Eq: potential condtion 2},
%Which means that we can set 
the ground state of the potential
leads to 
%\begin{align}\label{def varphi_0}
$\varphi\to\varphi_{0}$ $(\varphi'\to 0)$ asymptotically, 
%\end{align}
%without loss of generality \cite{Damour:2007uf},
resulting in 
\begin{align}\label{Eq:A0}
	A_0 = 1
\end{align}
As a result, \eqref{coupling func} can be used to make sure that the Jordan frame metric 
$g_{\mu \nu} \approx \eta_{\mu \nu}$ is asymptotically flat.

By following the discussions in Refs. \cite{Novak:1998rk, Sperhake:2017itk, Damour:1993hw}, 
 $\alpha_{0}$ ($\beta_{0}$) is positive (negative) to derive
the phenomena of the  spontaneous scalarization. 
The non-vanishing property of \eqref{Eq:E-frame requirement} implies that the parameter of
$\alpha$ in \eqref{Eq:alpha} can never be zero, 
i.e., the \emph{$\alpha$-constraint with}
\begin{align}\label{Eq:alpha-constraint}
	\alpha \neq 0,
\end{align}
which indicates that there exists an
\emph{unacceptable} critical value for $\varphi'$, denoted as 
%$\varphi'_{c}$, for $\varphi'$  such that $\alpha = 0$, i.e., 
\begin{align}
\varphi'_{c} \coloneqq -\frac{\alpha_0}{\beta_{0}}\,,
\end{align}
since this value leads to $\alpha = 0$ from
%due to the non-vanished value of 
\eqref{Eq:eq of alpha}.
Clearly, the derivative constraints in~\eqref{Eq:E-frame requirement}
are equivalent to the $\alpha$-constraint in~\eqref{Eq:alpha-constraint}.
Therefore, the forbidden value of $\varphi$ is
\begin{align}\label{new varphic}
	\varphi_{c} = \varphi'_{c} + \varphi_{0}.
\end{align}
This illustrates that the solution space of the scalar field $\varphi$
in the Einstein frame is divided
into two  branches by the value of $\varphi_{c}$.

%%%%%%%%%%%%%%%%%%%%%%%%%%%%%%%%%%%%%%%%%%%%%%%%%%%%%%%%%%%%%%
\section{Constraints on Specific Models}\label{Sec:Constraints on Specific Models}
\begin{figure}[t]
	\centering
 	\includegraphics[angle=90,scale=0.4]{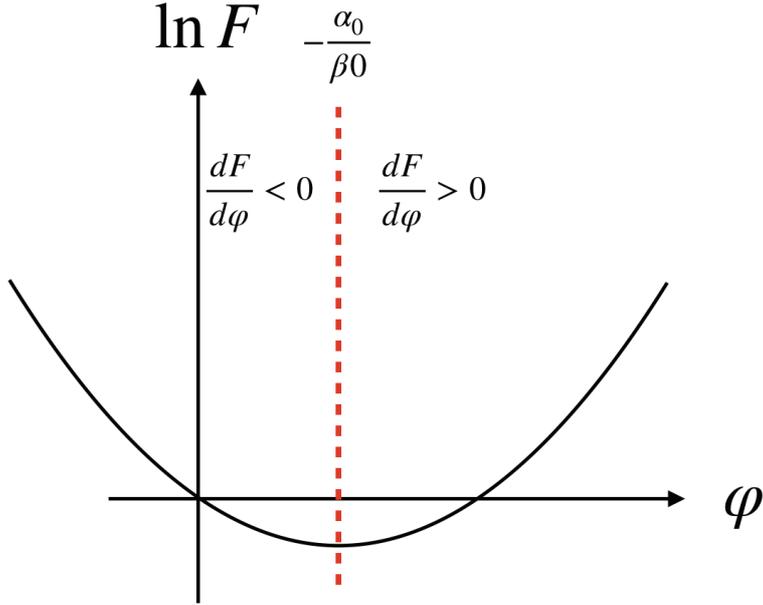}
	\caption{The parabola reveals the relation between $\ln F$ and $\varphi$, where
	         the dashed line represents the forbidden value $\varphi_c$ for $\varphi$.
	         Particularly, we have $F = \phi$ for the BDBW theory.}
	\label{Fig:parabola}
\end{figure}

In a specific model, the condition \eqref{Eq:E-frame requirement} would manifest itself as 
a critical value
$\varphi_{c}$. We will show that the scalar field can be nowhere equal 
to $\varphi_{c}$ as it cannot be a solution to the scalar equation in the Jordan frame.
Hence, the solution space of the scalar field in the Einstein frame is separated by $\varphi_{c}$ 
into two pieces. 
Moreover, in the PPN regime, we demonstrate that this critical value is determined by 
the coefficients of the coupling function \eqref{coupling func}.
In this section, we will take the general Jordan frame with the action \eqref{JD action} 
and BDBW theories as two examples for discussions. 
In particular, we concentrate on the potential in the Einstein frame 
with the form of 
\begin{align}\label{Eq:Eframe potential}
V(\varphi) = m^2\varphi^2,
\end{align}
which has  the minimum of $V_0=0$ at $\varphi_0=0$.
Hereafter, we drop out the superscript of prime on $\varphi '$ for simplicity 
due to $\varphi '=\varphi $. 
As a result, the parabola of \eqref{Eq:lnF} is shown as FIG.~\ref{Fig:parabola}.
%----------------- Figure 2 ---------------%
%------------------------------------------%

\subsection{General Jordan Frame Action}\label{Subsec:General Jordan Frame Action}
%The general Jordan frame with action \eqref{JD action} is discussed in the subsection.
%Subsequently, 
From \eqref{Eq:lnF} with $\varphi_0=0$, we get
% yield  
that
\begin{align}\label{varphi}
	\varphi = \varphi_{c} \pm \frac{\sqrt{\alpha_{0}^{2}-\beta_{0}\ln F}}{\beta_{0}}.
	%\coloneqq \varphi_{c} \pm \frac{\sqrt{\alpha_{0}^{2}-\beta_{0}\ln F}}{\beta_{0}}.
\end{align}
With the two signs in \eqref{varphi}, there are  different relations for $\varphi(\phi)$
or  $\phi(\varphi)$, resulting in different possible potentials of $U(\phi)$. 
Therefore, a given model in the Einstein frame may correspond to two
models in the Jordan frame.
%From \eqref{Eq:alpha} we have the relation
%\begin{align}\label{F_phi}
%F_{\phi} = F_{\varphi} \frac{d\varphi}{d\phi} = -2\alpha F \frac{d\varphi}{d\phi}.
%\end{align}

%\begin{align}
%\ln F = -2\alpha_{0}\varphi -\beta_{0}\varphi^{2}.
%\end{align}}

Moreover, the sign in \eqref{varphi} is associated with the behavior of $\varphi$.
We consider two situations: 
(a) $\varphi$ is \emph{not} always positive and (b) $\varphi$ is always positive.

%----------------- Figure 1 ---------------%
\begin{figure}[t]
	\centering
	\includegraphics[scale=0.6]{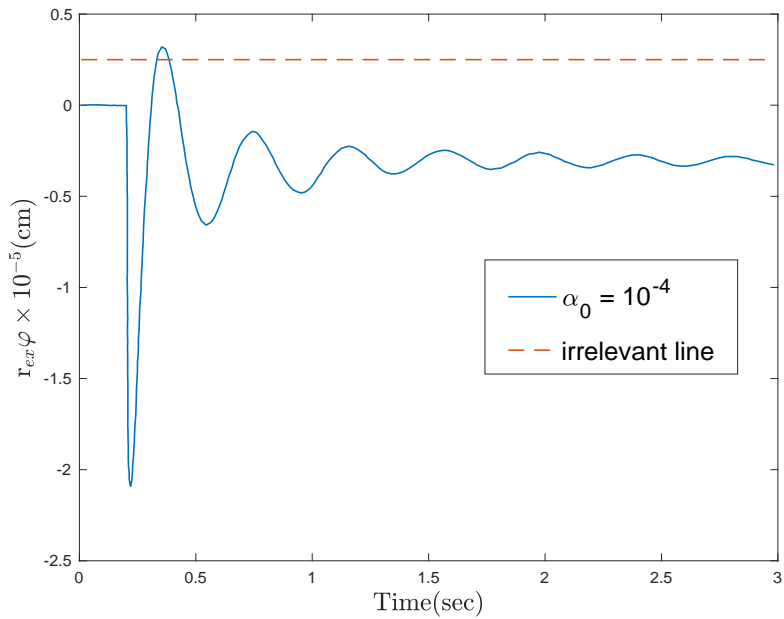}
	\caption{Waveforms of the scalar field, extracted at r$_{\text{ex}} = 5 \times 10^{9}$ cm
	  away from the supernovae core in the simulation of \cite{Sperhake:2017itk},
	  where the solid line represents that with $\alpha_{0}=10^{-4}$ in FIG.~1 of
	  \cite{Sperhake:2017itk} and the dashed horizontal line corresponds to the  
	  critical value of $\varphi_c  = 5 \times 10^{-6}$.}
	\label{Fig:varphi evolution}
\end{figure}
%------------------------------------------%
For (a), like all the solutions in~\cite{Sperhake:2017itk}, oscillating across
$\varphi =0$,  
we should choose the ``$+$'' sign in \eqref{varphi}, 
providing that $\varphi < \varphi_{c}$.\footnote{If 
we choose the ``$-$'' sign in \eqref{varphi}, 
$\varphi$ must be positive and greater than $\varphi_c$ due to $\beta_{0}<0$.}
Therefore, once the sign in \eqref{varphi} is determined, the other branch of the solution space 
represented by $\varphi > \varphi_{c}$ should collapse, 
so that $\varphi_{c}$ can be viewed as a ceiling of $\varphi$. 
In addition, we note that the collapsed region of $\varphi>\varphi_{c}$ is independent of the potential.
Any value of $\varphi$ exceeding this ceiling is not viable as it is not a solution to the scalar equation
in the Jordan frame associated with the current Einstein frame.
For example,   the evolution of the scalar field in the Einstein frame 
during the process of the supernovae explosion has been investigated in ~\cite{Sperhake:2017itk}. 
The solution labeled by $\alpha_{0}=10^{-4}$ violates this constraint 
as illustrated in FIG.~\ref{Fig:varphi evolution}. 

Although the solution space is constrained by $\varphi < \varphi_{c}$, 
 we still cannot specify a unique model in the Jordan frame due to the lack of the information 
about $F_{\phi}$ to determine the form of the relation $\phi(\varphi)$.
However, we can constrain $\alpha$ to be positive as $\varphi<\varphi_{c}$,
which gives rise to
\begin{align}
	F_{\phi} \frac{d\phi}{d\varphi} < 0
\end{align}
due to \eqref{Eq:alpha}.
%In spite of the sign of~\eqref{trans relation}  underdetermined 
%without a further information of $F_{\phi}$,
Clearly, the positivity of $F_{\phi}$ implies that
the sign of~\eqref{trans relation} should be minus and vice versa. 
Furthermore, we have two models in the Jordan frame based on  the ``$\pm$'' signs of $F_{\phi}$,
within  $\varphi<\varphi_{c}$.
For (b), there are two cases. The first one corresponds to
the ``$-$'' sign  in~\eqref{varphi} with $\varphi > \varphi_{c}$.
Similar to the argument above, 
there exists two models in the Jordan frame related to the sign of $F_{\phi}$. 
In this case with the equivalence of $\alpha < 0$, we have the relation 
\begin{align}
	F_{\phi}  \frac{d\phi}{d\varphi} > 0.
\end{align}
By contrast, the case for the ``$+$'' sign requires $0 < \varphi < \varphi_{c}$.
The solution space is bounded, which is the most restrictive one among the cases.

By differentiating both sides of \eqref{varphi} with respect to $\phi$, 
one gets an additional equation
\begin{align}\label{dvarphi/dphi}
\frac{d\varphi}{d\phi} = \frac{\mp F_{\phi}}{2F\sqrt{\alpha_{0}^{2}-\beta_{0}\ln F}},
\end{align}
which suggests  that $F_{\phi}$ determines the sign of $d\varphi/d\phi$ since $F>0$. 
We note that the sign in \eqref{dvarphi/dphi} is opposite to the one chosen in \eqref{varphi}.
The requirement of $\ln F > \alpha_{0}^{2}/\beta_{0}$ 
to prevent $\varphi$ from being imaginary is fulfilled
because \eqref{Eq:lnF} has minimum $\alpha_0^2/\beta_0$ at the forbidden point $\varphi_c$,
which can easily be checked from  FIG.~\ref{Fig:parabola}.
One can see  that the discussions of the sign problem of $F_{\phi}$ above 
are consistent with the sign in \eqref{dvarphi/dphi} .

%%%%%%%%%%%%%%%%%%%%%%%%%%%%%%%%%%%%%%%%%%%%%%%%%%%%%%%%%%%%%%
\subsection{Brans-Dicke-Bergmann-Wagoner theory}\label{Subsec:BDBW}
Another example is the BDBW theory with $F = \phi$ in the action \eqref{JD action}, particularly
given by
\begin{align}\label{case 2 action}
S = \int d^{4}x\, \frac{\sqrt{-g}}{16\pi G} \bigg( \phi R - \frac{\omega(\phi)}{\phi} 
g^{\mu \nu} \partial_{\mu}\phi \partial_{\nu}\phi -U(\phi)\bigg) +S_{m}[\psi_{m}, g^{\mu \nu}].
\end{align}
It can be found that the sign of $d\varphi/d\phi$ can totally be determined 
through  $\alpha$ due to $F_{\phi} = 1$.
In this subsection, we  will transform  the potential defined 
in the Einstein frame back into the Jordan one and discuss  
the ambiguity arising from this transformation.
In this specific case, since $F(\phi) = A^{-2}$ is simply $\phi$,  
we have 
\begin{align} \label{phi>0}
\phi = A^{-2} >0,
\end{align}
leading to 
\begin{align}\label{ln phi}
	\ln \phi = -2\alpha_{0}\varphi - \beta_{0}\varphi^{2}\,,
\end{align}
which is also illustrated in FIG.~\ref{Fig:parabola}. 

From FIG.~\ref{Fig:parabola}, it is easy to see that the condition
\eqref{Eq:E-frame requirement} is equivalent to
$\varphi \ne \varphi_{c}$.
As a result, one finds that \eqref{Eq:eq of alpha} can never be zero. Consequently, 
\eqref{Eq:alpha} reads as 
\begin{align}\label{Eq:alpha for BDBW}
\alpha = -\frac{1}{2\phi}\frac{d\phi}{d\varphi}.
\end{align}
From \eqref{Eq:alpha for BDBW}, we obtain
the inequality by \eqref{phi>0}
\begin{align}\label{alpha dvarphi/dphi}
	\alpha\frac{d\varphi}{d\phi}  = -\frac{1}{2\phi} <0.
\end{align}
As mentioned early that $d\varphi/d\phi$ is either positive or negative,  
the valid value of $\varphi$ can fix the sign 
in \eqref{trans relation} as follows.
For $\varphi > \varphi_{c}$, we have $\alpha<0$, so that one has
\begin{align}
	\frac{d\varphi}{d\phi}
%	= \sqrt{\frac{3(F_{,\phi})^{2}}{4F^{2}}+\frac{\omega}{2\phi F}}
	= \frac{\sqrt{3+2\omega}}{2\phi}
\end{align}
by \eqref{alpha dvarphi/dphi}. 
Similarly, the redefinition of the scalar fields leads to 
\begin{align} 
	\frac{d\varphi}{d\phi}= -\frac{\sqrt{3+2\omega}}{2\phi}
\end{align}
for $\varphi < \varphi_{c}$.

In the limit of the BD theory, $\omega$ in \eqref{case 2 action} is constant, i.e.,
$\omega(\phi) = \omega_{\text{BD}}$.
Consequently, the equation in \eqref{Eq:alpha for BDBW} reveals that
\begin{align}\label{Eq:alpha for BD}
	 \alpha = \alpha_{0}+\beta_{0}\varphi = \mp \frac{1}{\sqrt{3+2\omega_{\text{BD}}}}
\end{align} 
where the minus (plus) case corresponds to 
the branch of $\varphi > \varphi_{c}$ ($\varphi < \varphi_{c}$).
Since the right-hand side of \eqref{Eq:alpha for BD} is constant, 
we have two possibilities: 
(i) $\beta_{0}=0$ and (ii) $\varphi$ is almost a constant. 
Due to the negativity of $\beta_{0}$ in this paper, 
the solution is restricted to the second one.

It is interesting to note that $\omega_{\text{BD}} \to \infty$ is a critical value, 
equivalent to $\varphi =\varphi_{c}$ due to $\alpha = 0$.
Even though this case is consistent with those in the literature and in turn reproduces 
the results in GR in the sense that all its predictions become indistinguishable 
from GR in the \emph{Jordan frame} \cite{Will:2014kxa},
it cannot be transformed to the \emph{Einstein frame} due to the violation of 
the derivative constraints and $\alpha$-constraint.
Therefore, the case of the infinite $\omega_{\text{BD}}$ is improper to be discussed in
the \emph{Einstein frame}, which shows the \emph{inequivalence} between the two frames.

% and hence is out of the regime under our consideration. 

We again note that the critical value of $\varphi_{c}$ is so generic that the results of all the simulations
should obey the criterion $\varphi \ne \varphi_{c}$. 
For instance, it has been concentrated on the BDBW theory to investigate the 
dynamical scalarization of the neutron star binaries in ST in \cite{Palenzuela:2013hsa}. 
The critical value for the scalar field in their setting is $\varphi_{c} = 0$, 
and the results therein do all satisfy this criterion, i.e., 
no crossing the line of $\varphi =0$.

%To explicitly write down the action in the Jordan frame, we 
Starting from  \eqref{ln phi} in the BDBW theory, we get
\begin{align}\label{Eq: varphi of phi BD}
	\varphi = \varphi_{c} \pm \frac{\sqrt{\alpha_{0}^{2}-\beta_{0}\ln\phi}}{\beta_{0}}.
\end{align}
Since $U = 4VA^{-4} =4m^{2} \varphi^{2} \phi^{2}$ as defined in~\eqref{Eq:Einstein action},
we  recover the potential $U$ in the  Jordan frame, given by
\begin{align}\label{Eq:U(phi)}
U(\phi) = 4m^{2}\beta_{0}^{-2}
\bigg( 2\alpha_{0}^{2}-\beta_{0}\ln\phi \mp 2\alpha_{0}\sqrt{\alpha_{0}^{2}-\beta_{0}\ln\phi} \bigg)\phi^{2}
\end{align}
and
\begin{align}\label{Eq:U(varphi)}
U(\varphi) = 4m^{2}\varphi^2\exp\big(-4\alpha_{0}\varphi-2\beta_{0}\varphi^2\big),
\end{align}
 in terms of $\phi$ and $\varphi$, respectively,
indicating two different potentials in the Jordan frame, 
which depend on the
branches of the scalar field  in the Einstein frame, 
i.e., the sign chosen in~\eqref{Eq: varphi of phi BD}.
For the case of the action \eqref{JD action} in Sec.~\ref{Subsec:General Jordan Frame Action},
one cannot consider only the ``$+$'' sign 
in~\eqref{trans relation}, which is equivalent to choose one model from two prospective ones.

In addition, the undetermined signs in~\eqref{Eq: varphi of phi BD} can be eliminated 
by considering the function of $(\varphi - \varphi_c)^2$.
Hence, the solution space has an $\mathbb{Z}_{2}$ symmetry 
with respect to $\varphi_c$.
In other words,
 the model in the Jordan frame, which is responsible for the given model in
the Einstein frame, remains the same under the transformation 
$\varphi -\varphi_{c} \longleftrightarrow -(\varphi -\varphi_{c})$. 
We can shift $\varphi$ to $\bar{\varphi} = \varphi - \varphi_{c}$,
resulting in that the existence of the $\mathbb{Z}_{2}$ symmetry is characterized by an even 
 potential $U(\bar{\varphi})$  of $\bar{\varphi}$.
%However, t
The potential \eqref{Eq:U(varphi)}, which can be rewritten as
\begin{align}\label{Eq:U(bar varphi)}
U(\bar{\varphi}) = 4m^{2}(\bar{\varphi} + \varphi_{c})^2
\exp\Big[-2\beta_{0}\big(\bar{\varphi}^2- \varphi_{c}^2\big)\Big],
\end{align}
 is not a even function of $\bar{\varphi}$ apparently.
As a result, the one to one correspondence between models in both frames holds strictly.
In general, it is hard to have the same symmetry of $\mathbb{Z}_{2}$ 
at $\varphi_c$ for  $U(\varphi)$.
Particularly, the direct effect of the $\mathbb{Z}_{2}$ symmetry 
at $\varphi_c$ for the potential is that the model in the Einstein frame is 
associated with a unique model in the Jordan frame. 

%%%%%%%%%%%%%%%%%%%%%%%%%%%%%%%%%%%%%%%%%%%%%%%%%%%%%%%%%%%%%%
\section{Conclusions}
In order to use the formulation of the ST theories in the Einstein frame,
we have investigated the redefinition of the scalar.
% in the ST theory.
We have found that the regularity of such redefinition~\eqref{trans relation},
which comes from the requirement $d\varphi/d\phi\neq0$ and 
has been shown to be equivalent to the condition $\alpha\neq0$,
gives an irrelevant value $\varphi_c$  of  the scalar field in the Einstein frame.
The value of $\varphi_c$ separates the solution space of the scalar field into two viable regions.
Any scalar field in the Einstein frame, which crosses the irrelevant value,  
cannot be the solution in the Jordan frame.

The two signs in~\eqref{trans relation} result in two different models 
in the Jordan frame. 
%Thus a model in the Einstein frame usually corresponds
%to two models in the Jordan frame while i
In some special  cases, such as that in the BDBW theory without a potential,
these two models in the Jordan frame coincide with each other.
In general, the sign of~\eqref{trans relation} can be obtained by  $F_{\phi}$
and $\varphi$. For the case with  one branch of the solution
space collapsed, the sign of~\eqref{trans relation} can be fixed by the sign of $F_{\phi}$
alone. In such a case, it is the ambiguity of the sign of $F_{\phi}$ that gives rise to 
two prospective models in the Jordan frame.
On the other hand, in some models with $F_{\phi}$ being constant, such as the BDBW theory,
the sign of~\eqref{trans relation} is fully determined by the branch of the scalar fields.
%which is discussed in Sec.~\ref{Subsec:BDBW}.
If we consider the asymptotic value $\varphi_{0} \neq 0$,
\eqref{varphi} should be modified as
\begin{align}\label{new varphi of phi}
	\varphi -\varphi_{0} = \varphi'_{c}\pm \frac{\sqrt{\alpha_{0}^{2}-\beta_{0}\ln F}}{\beta_{0}}.
\end{align}
Together with \eqref{new varphic}, one can also conclude that
if the potential $U(\varphi)$ in the Jordan frame is an even function 
with respect to the value of $\varphi = \varphi_{c}$, 
 the correspondence between the models in two frames is one to one. 
Hence, by satisfying 
the $\alpha$-constraint in \eqref{Eq:alpha-constraint} 
the results in the Einstein frame can be well-defined under this condition.

We have demonstrated that
the critical value $\varphi_{c}$ provides a viable constraint 
on the formulation of the ST theories in the Einstein frame,
which is even independent of the form of the potential. Clearly,  all the numerical  results
must obey the criterion of no crossing the line of the critical value. Otherwise, they will
not have any physical meaning in the Einstein frame.
In addition, we have also shown that the case of the
infinite $\omega_{\text{BD}}$ in the BD theory in the Einstein frame 
is irrelevant due to the violation of the derivative constraints and $\alpha$-constraint.
%,which results in the \emph{inequivalence} between two frames.

Furthermore, for the case without a potential, 
two branches of the solution space of the scalar field in the Einstein frame correspond to
the same model in the Jordan one. 
However, the solution branch of $\varphi$ should be specified 
because the $\alpha$-constraint is strict with the behavior of $\varphi$
during the evolution in the Einstein frame.

Finally, we conclude that the critical value $\varphi_{c}$ induced from 
the $\alpha$-constraint or derivative constraints
and the sign determination in \eqref{trans relation} are two important issues
to study the ST theory in the Einstein frame.

%%%%%%%%%%%%%%%%%%%%%%%%%%%%%%%%%%%%%%%%%%%%%%%%%%%%%%%%%%%%%%
\begin{acknowledgments}
	We would like to thank Patrick Chi-Kit Cheong for useful discussions.
	The work was partially supported by National Center for Theoretical Sciences,
	Ministry of Science and Technology (MoST-107-2119-M-007-013-MY3 
	and MoST-108-2811-M-001-598), and
	% LWL is supported by 
	Academia Sinica Career Development Award Program (AS-CDA-105-M06).
\end{acknowledgments}

%%%%%%%%%%%%%%%%%%%%%%%%%%%%%%%%%%%%%%%%%%%%%%%%%%%%%%%%%%%%%%

\end{document}